\documentclass[usenatbib]{mn2e}
\usepackage{epsfig}

\newcommand{\mnras}{MNRAS}
\newcommand{\apj}{ApJ}
\newcommand{\apjl}{ApJL}
\newcommand{\apjs}{ApJS}
\newcommand{\prd}{Phys.~Rev.~D}
\newcommand{\aj}{AJ}
\newcommand{\araa}{ARA\&A}

\def\gsim { \lower .75ex \hbox{$\sim$} \llap{\raise .27ex \hbox{$>$}} }
\def\lsim { \lower .75ex \hbox{$\sim$} \llap{\raise .27ex \hbox{$<$}} }
\begin{document}
\title[Evolution of Halo Mass Profiles] {The redshift
dependence of the structure of massive $\Lambda$CDM halos}
\author[Gao et al.] {
\parbox{\textwidth}{Liang Gao$^1$, Julio
F. Navarro$^{2,3}$\thanks{Fellow of the Canadian Institute for
Advanced Research}, Shaun Cole$^1$, Carlos S. Frenk$^1$, Simon
D.M. White$^4$, Volker Springel$^4$, Adrian Jenkins$^1$, Angelo F. Neto$^{5}$}
\vspace*{4pt} \\
$^1$Institute of Computational Cosmology, Department of Physics,
University of Durham,\\ Science Laboratories, South Road, Durham DH1
3LE, UK \\ 
$^2$Astronomy Department, University of Massachusetts, LGRT-B 619E,
Amherst, MA 01003, USA \\ 
$^3$Department of Physics and Astronomy, University of Victoria, PO
Box 3055 STN CSC, Victoria, BC, V8W 3P6 Canada \\ 
$^4$Max-Planck Institute for Astrophysics, Karl-Schwarzschild Str. 1,
D-85748, Garching, Germany \\
$^5$Instituto de F\'{i}sica, Universidade Federal do Rio Grande do
Sul, Porto Alegre RS, Brazil \\ 
}
\maketitle

\begin{abstract}
We use two very large cosmological simulations to study how the density
profiles of relaxed $\Lambda$CDM dark halos depend on redshift and on halo
mass. We confirm that these profiles deviate slightly but systematically from
the NFW form and are better approximated by the empirical formula, $d\log
\rho/d\log r \propto r^{\alpha}$, first used by Einasto to fit star counts in
the Milky Way. The best-fit value of the additional shape parameter, $\alpha$,
increases gradually with mass, from $\alpha\sim 0.16$ for present-day galaxy
halos to $\alpha\sim 0.3$ for the rarest and most massive clusters. Halo
concentrations depend only weakly on mass at $z=0$, and this dependence
weakens further at earlier times. At $z\sim 3$ the average concentration of
relaxed halos does not vary appreciably over the mass range accessible to our
simulations ($M \, \gsim \, 3\times 10^{11} \, h^{-1} M_\odot$). Furthermore,
in our biggest simulation, the average concentration of the {\em most
massive}, relaxed halos is constant at $\langle c_{200}\rangle \sim 3.5$ to 4
for $0\leq z\leq 3$.  These results agree well with those of Zhao et al
(2003b) and support the idea that halo densities reflect the density of the
universe at the time they formed, as proposed by Navarro, Frenk \& White
(1997). With their original parameters, the NFW prescription overpredicts halo
concentrations at high redshift. This shortcoming can be reduced by
modifying the definition of halo formation time, although the evolution
of the concentrations of Milky Way mass halos is still not reproduced well.
In contrast, the much-used revisions of the NFW prescription by Bullock et
al. (2001) and Eke, Navarro \& Steinmetz (2001) predict a steeper drop in
concentration at the highest masses and stronger evolution with redshift than
are compatible with our numerical data. Modifying the parameters of
these models can reduce the discrepancy at high masses, but the overly rapid
redshift evolution remains.  These results have important implications for
currently planned surveys of distant clusters.
\end{abstract}

\begin{keywords}
methods: N-body simulations -- methods: numerical --dark matter --
galaxies: haloes -- galaxies:structure
\end{keywords}

\section{Introduction}
Over the past decade, cosmological N-body simulations have shown consistently
that equilibrium dark matter halos have spherically-averaged mass density
profiles which are approximately ``universal'' in form; i.e., their shape is
independent of mass, of the values of the cosmological parameters, and of the
linear power spectrum from which nonlinear structures have grown. As a result,
it is useful to parametrize halo profiles by simple empirical formulae, such
as that proposed by Navarro, Frenk \& White (1995, 1996, 1997, hereafter NFW):
\begin{equation}
\frac{\rho(r)}{\rho_{\rm crit}}= \frac{\delta_{\rm c} }{(r/r_{\rm
s})(1+r/r_{\rm s})^2},
\label{eq:nfw}
\end{equation}
where $\rho_{\rm crit}=3H^2/8\pi G$ is the critical density for
closure{\footnote{We express Hubble's constant as $H(z)$ and its
present-day value as $H(z=0)=H_0=100 \, h$ km s$^{-1}$
Mpc$^{-1}$.\label{ftn:hz}}}, $\delta_c$ is a characteristic density
contrast, and $r_s$ is a scale radius. Note that this formula contains
two scale parameters but no adjustable shape parameter.

As discussed in some detail by NFW and confirmed by subsequent numerical work,
the two parameters of the NFW profile do not take arbitrary values, but are
instead correlated in a way that reflects the mass-dependence of halo assembly
times (e.g., Kravtsov, Klypin \& Khokhlov 1997; Avila-Reese et al.  1999; Jing
2000; Ghigna et al. 2000; Bullock et al. 2001; Eke, Navarro \& Steinmetz 2001;
Klypin et al. 2001). The basic idea behind this interpretation is that the
characteristic density of a halo tracks the mean density of the universe at the
time of its formation.  Thus, the later a halo is assembled, the lower its
characteristic density, $\delta_c$, or, equivalently, its ``concentration'' (see
Section ~\ref{ssec:hcat} for a definition).

Although the general validity of these trends is well established, a
definitive account of the redshift and mass dependence of halo concentration
is still lacking, even for the current concordance cosmology. This is
especially true at high masses, where enormous simulation volumes are required
in order to collect statistically significant samples of these rare
systems. Simulating large cosmological volumes with good mass resolution is a
major computational challenge, and until recently our understanding of the
mass profile of massive halos has been rather limited, derived largely from
small numbers of individual realizations or from extrapolation of
models calibrated on different mass scales (NFW; Moore et al. 1998; Ghigna et
al. 2000; Klypin et al. 2001; Navarro et al. 2004; Tasitsiomi et al. 2004;
Diemand et al. 2004; Reed et al. 2005).

Individual halo simulations may result in biased concentration estimates,
depending on the specific selection criteria used to set them up.  In
addition, they are unlikely to capture the full scatter resulting from the
rich variety of possible halo formation histories.  Extrapolation based on
poorly tested models can also produce substantial errors, as recently
demonstrated by Neto et al. (2007). These authors analyzed the 
mass-concentration relation for halos identified at $z=0$ in the {\it
Millennium Simulation} of Springel et al. (2005, hereafter {\tt MS}) and
confirmed the earlier conclusion of Zhao et al. (2003b) that the
models of Bullock et al. (2001, hereafter B01) and Eke et al. (2001,
hereafter ENS) (which were calibrated to match galaxy-sized halos) severely
underestimate the average concentration of massive clusters, by up to a factor
of $\sim 3$.

Estimates of concentrations can also be biased by the inclusion of
unrelaxed halos. These often have irregular density profiles caused by major
substructures. Smooth density profiles are often poor fits to such halos,
and the resulting concentration estimates are ill-defined because they
depend on the radial range of the fit and choice of weighting. They can also
lead to spurious correlations (see for example figure~9 of Neto et
al. 2007). Consequently, in this paper we follow Neto et al. and select only
relaxed halos for analysis. This is not without its own problems. Such
selection biases against recently formed halos, which may preferentially
have lower concentrations. However we believe this is preferable to
polluting the sample with meaningless concentration estimates of the kind
that arise when smooth spherical models are force-fit to lumpy, multi-modal
mass distributions. Hayashi \& White (2008) stacked {\it all} halos,
regardless of dynamical state, in the {\tt MS} and studied the resulting
mean profiles as a function of halo mass. The relatively small differences
between their results and those found below shows that the inclusion of
unrelaxed halos has rather little effect in the mean. 

A further preoccupation concerns indications that halo profiles deviate
slightly but systematically from the NFW model (Moore et al. 1998, Jing \& Suto
2000, Fukushige \& Makino 2001, Navarro et al. 2004, Prada et al. 2006, Merritt
et al. 2006), raising the possibility that estimating concentrations by
force-fitting simple formulae to numerical data may result in subtle biases
that could mask the real trends. This is especially important because of hints
that such deviations depend systematically on halo mass (Navarro et al.
2004, Merritt et al. 2005). Evaluating and correcting for such deviations is
important in order to establish conclusively the mass and redshift dependence
of halo concentration.

These uncertainties are unfortunate since observations, especially at high
redshift, often focus on exceptional systems. For example, massive galaxy
clusters are readily identified in large-scale surveys of the distant
universe, and understanding their internal structure will be critical for the
correct interpretation of cluster surveys intending to constrain the nature of
dark energy. These will make precise measurements of the evolution of cluster
abundance in samples detected by gravitational lensing, by their optical or
X-ray emission, or through the Sunyaev-Zel'dovich effect (see, e.g., Carlstrom
et al. 2002, Hu 2003, Majumdar \& Mohr 2003, Holder 2006, and references
therein).

There is at present no {\it ab initio} theory that can reliably predict the
internal structure of CDM halos. The models of Zhao et al (2003a),
Wechsler et al (2002) (see also Lu et al. 2006) are interesting, but as shown
in Neto et al (2007) they account for only a small fraction in the measured
dispersion in concentration at a fixed mass. There is currently no
substitute for direct numerical simulation when detailed predictions are
needed for comparison with observation.

We address these issues here by combining results from the {\tt MS} with
results from an additional simulation which followed a substantially smaller
volume but with better mass resolution.  This allows us to extend the range of
halo masses for which we can measure concentrations and to assess how these
measures are affected by numerical resolution. Our analysis procedure follows
closely that of Neto et al. (2007). In particular, we concentrate in this paper
on the properties of halos which are relaxed according to the criteria defined
by these authors; mean density profiles for {\it all} {\tt MS} halos of given
mass, regardless of dynamical state, are presented by Hayashi \& White
(2008). We begin in Section~\ref{sec:numexp} by describing briefly the
numerical simulations and the halo catalogue on which this study is based. In
Section~\ref{sec:res} we present our main results for the dependence of profile
shape and concentration on halo mass and redshift. We conclude with a brief
discussion and summary in Section~\ref{sec:conc}.

\begin{figure*} \hspace{-0.5cm}
\resizebox{8cm}{!}{\includegraphics{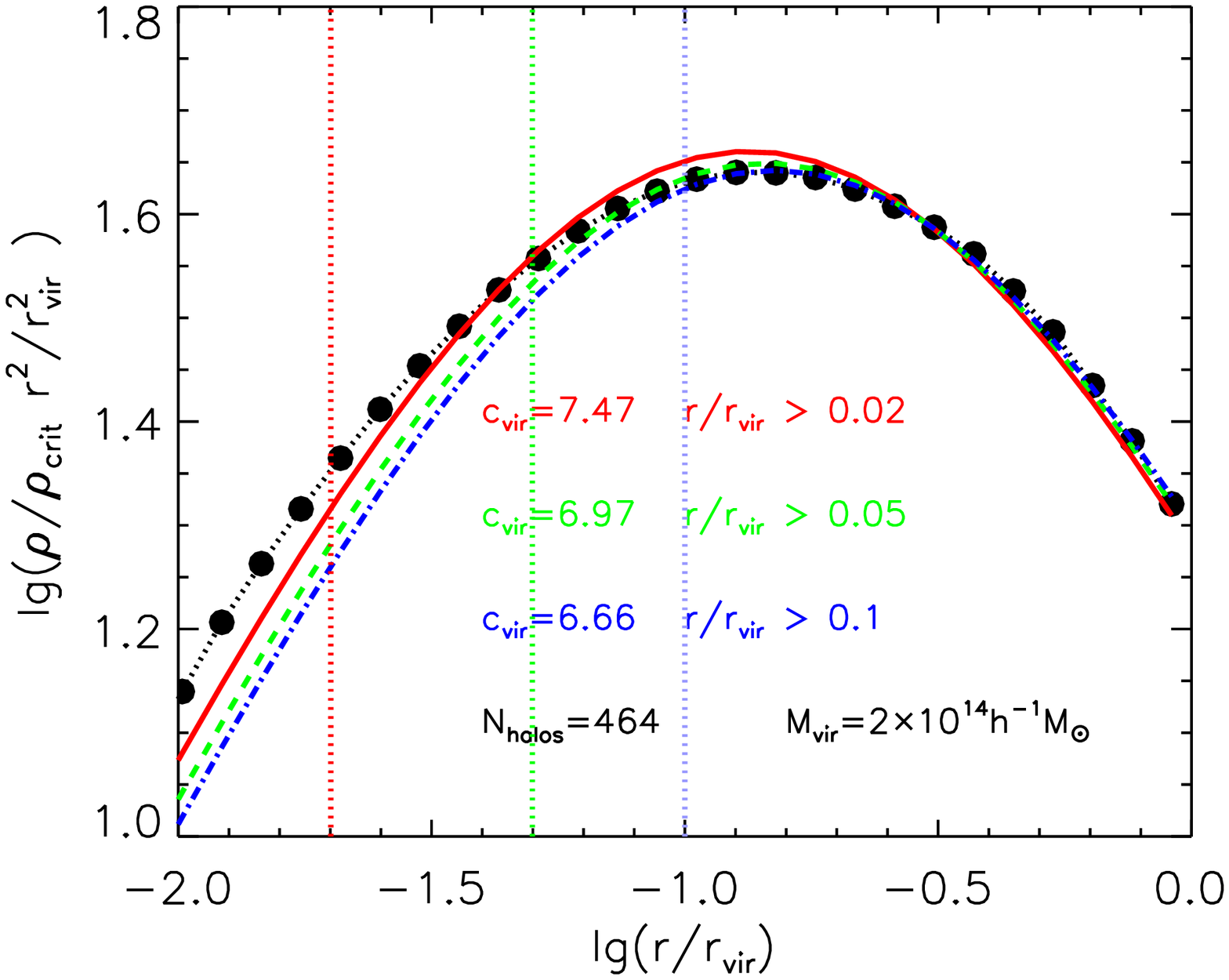}}
\resizebox{8cm}{!}{\includegraphics{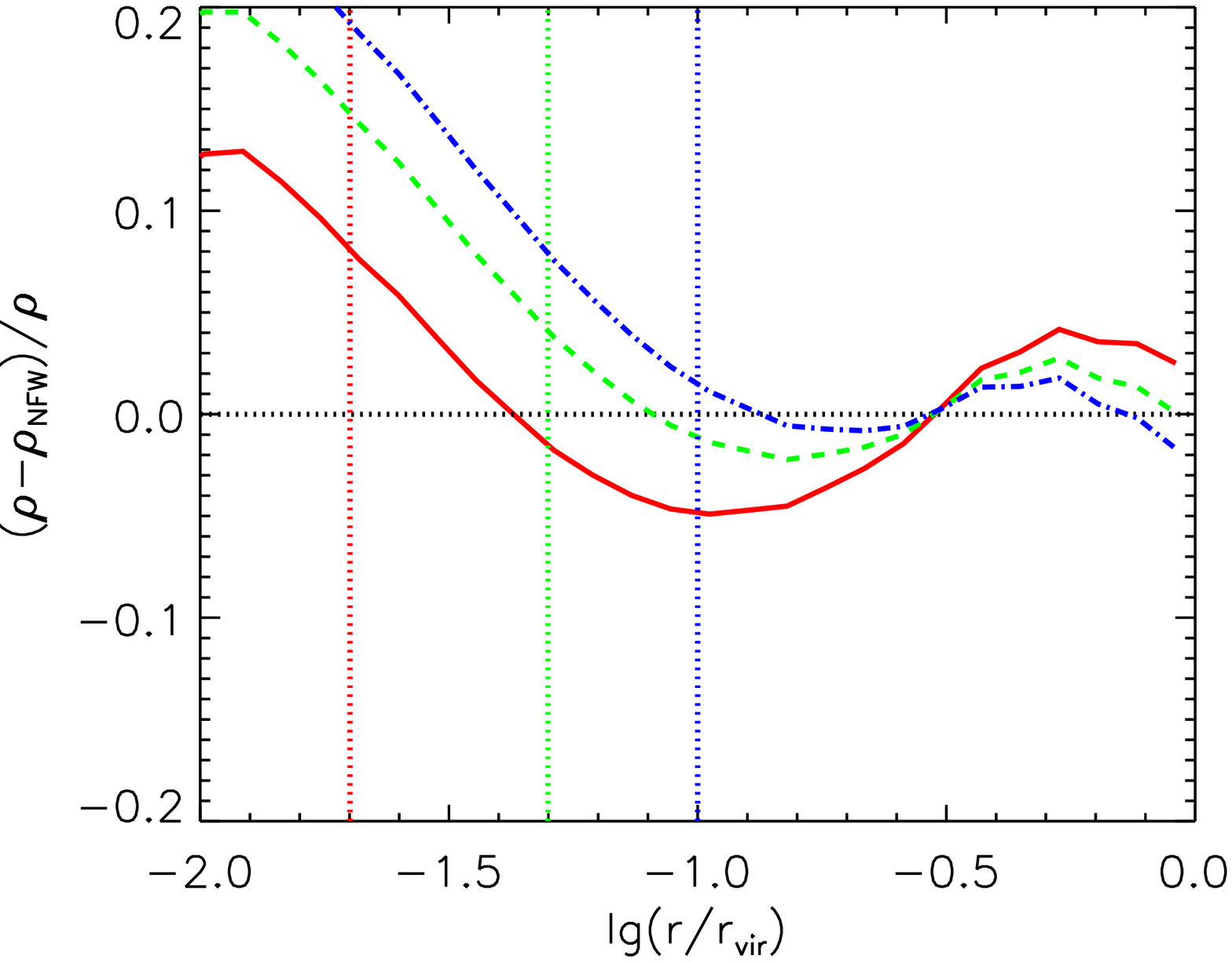}}
\resizebox{8cm}{!}{\includegraphics{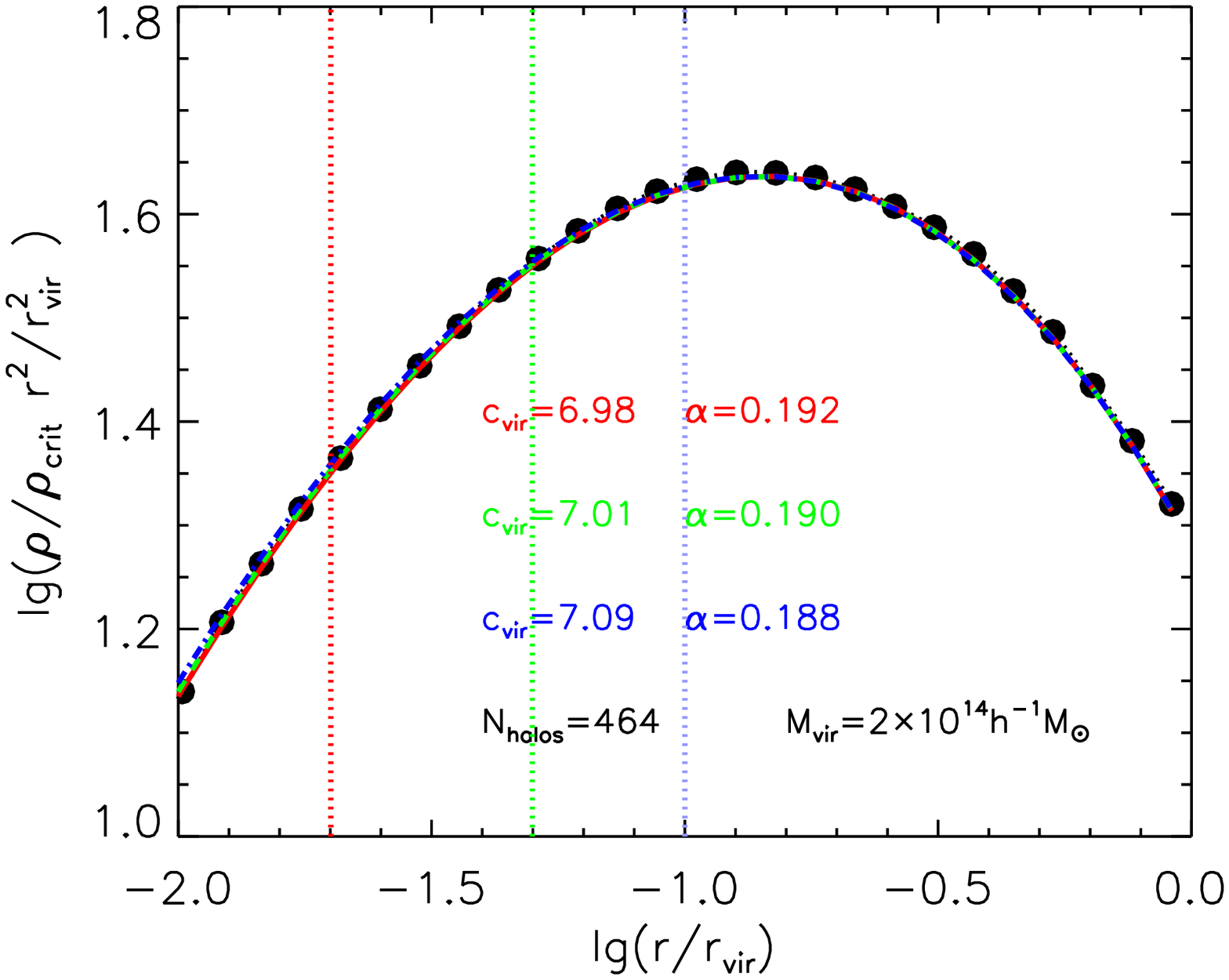}}
\resizebox{8cm}{!}{\includegraphics{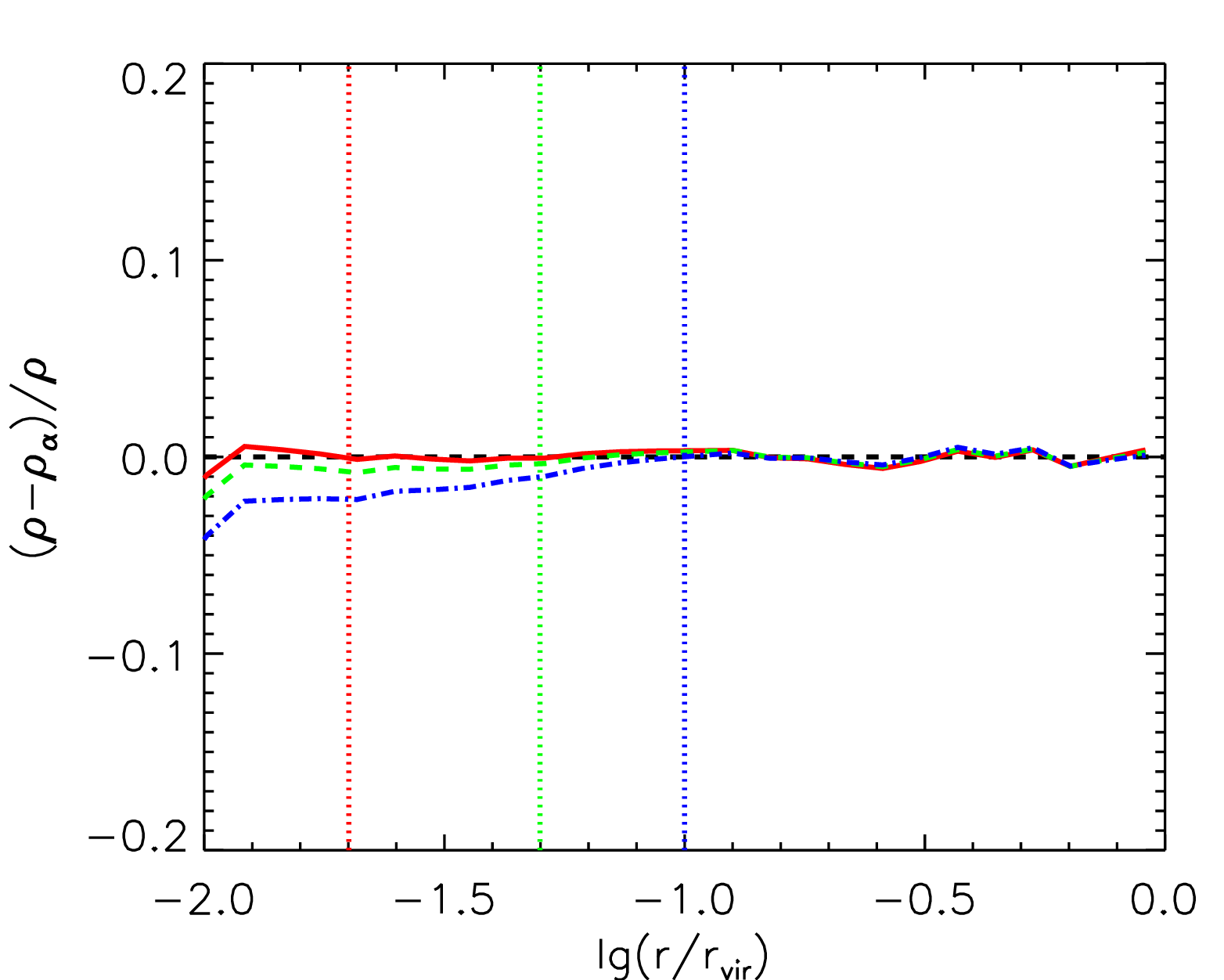}}
\caption{{\it Left panels:} The stacked density profile of 464 halos of virial
mass $ \sim 2 \times 10^{14}h^{-1}M_\odot$, identified at $z=0$ in the {\tt
MS}. The curves in the upper left panel show different NFW fits obtained by
varying the radial range fitted: $[0.02,1]r_{\rm vir}$ (solid red),
$[0.05,1]r_{\rm vir}$ (dashed green) and $[0.1,1]r_{\rm vir}$ (dot-dashed
blue).  Note the small disagreement between the actual profile shape and the
NFW model (upper panels). This leads to concentration estimates which depend
slightly on the innermost radius of the fit. Fits using
eq.~(\ref{eq:rhoalpha}) are more robust to such variations in fitting range,
as shown in the lower left panel. Panels on the right show the residuals
corresponding to the fits shown on the left.}
\label{fig:r2rho}
\end{figure*}

\section{The Numerical Simulations}
\label{sec:numexp}

The analysis presented here is based primarily on halos identified in the {\it
Millennium Simulation} ({\tt MS}) of Springel et al. (2005). The halo
identification and cataloguing procedure follows closely that described in
detail by Neto et al. (2007). For completeness, we here recapitulate the main
aspects of the procedure, referring the interested reader to the earlier
papers for details.

\subsection{Simulations}

The {\tt MS} is a large N-body simulation of the concordance $\Lambda$CDM
cosmogony. It follows $N=2160^3$ particles in a periodic box of side $L_{\rm
box}=500 \, h^{-1}{\rm Mpc}$. The chosen cosmological parameters were
$\Omega_{\rm m}=\Omega_{\rm dm}+\Omega_{\rm b}=0.25$, $\Omega_{\rm b}=0.045$,
$h=0.73$, $\Omega_{\rm \Lambda} = 0.75$, $n=1$, and $\sigma_8=0.9$. Here
$\Omega$ denotes the present-day contribution of each component to the
matter-energy density of the Universe, expressed in units of the critical
density for closure, $\rho_{\rm crit}$; $n$ is the spectral index of the
primordial density fluctuations, and $\sigma_8$ is the rms linear mass
fluctuation in a sphere of radius $8 \, h^{-1}$ Mpc at $z=0$. The particle
mass in the {\tt MS} is $8.6 \times 10^8 \, h^{-1}{\rm M_{\odot}}$. Particle
interactions are softened on scales smaller than the (Plummer-equivalent)
softening length, $\epsilon=5\, h^{-1}{\rm kpc}$.

We also use a second simulation of a smaller volume ($100^3 \, h^{-3}$
Mpc$^3$) within the same cosmological model. This simulation followed
$N=900^3$ particles of mass $9.5 \times 10^7\, h^{-1}{\rm M_{\odot}}$ and
softened interactions on scales smaller than $\epsilon=2.4\, h^{-1}{\rm
kpc}$. It thus has approximately 9 times better mass resolution than the {\tt
MS}. Hereafter we refer to it as the {\tt hMS}.

\begin{figure*} \hspace{-0.5cm}
\resizebox{8cm}{!}{\includegraphics{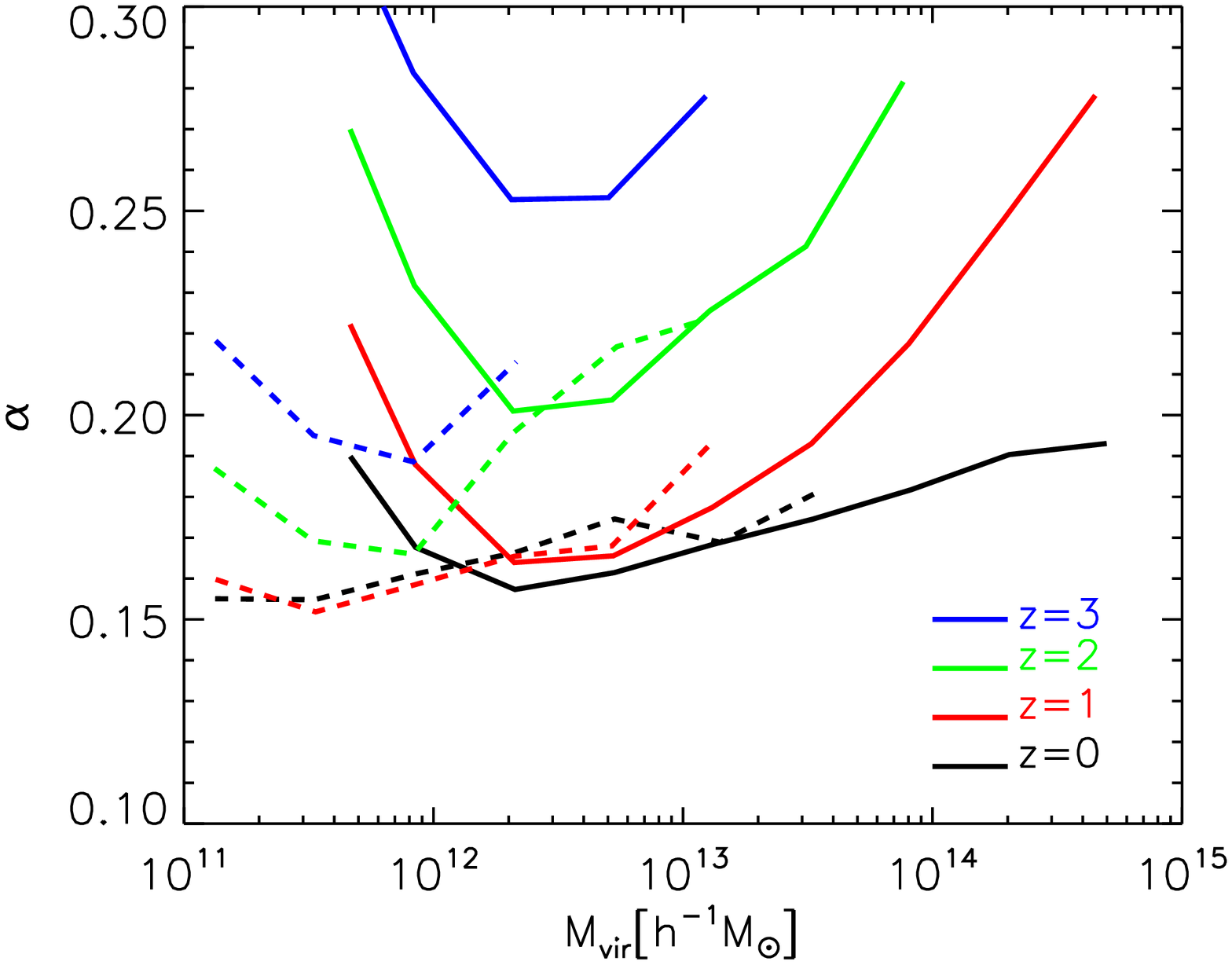}}
\resizebox{8cm}{!}{\includegraphics{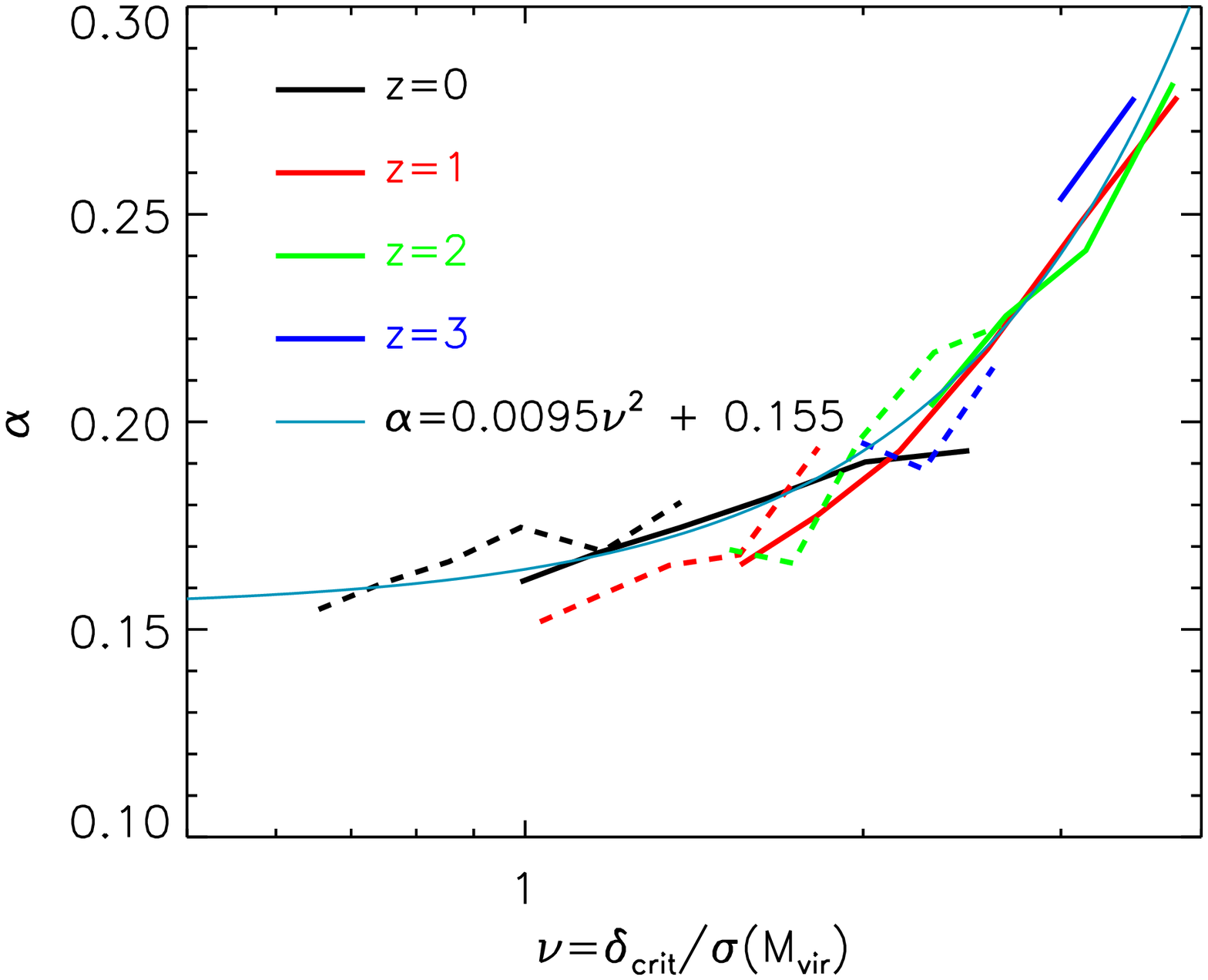}}
\caption{{\it Left panel:} The best-fit shape parameter, $\alpha$
(eq.~(\ref{eq:rhoalpha})), as a function of halo mass and redshift, after
binning and stacking halos by mass. Solid and dashed lines correspond to the
{\tt MS} and {\tt hMS} simulations, respectively. Only results corresponding
to halos with more than $\sim 500$ particles and to stacks with more than $10$
halos are shown. Note the good general agreement between the two simulations.
Differences are substantial only where the number of particles is less than
$3000$. {\it Right panel:} Values of $\alpha$ for halos with more than $3000$
particles plotted as a function of the dimensionless ``peak height''
parameter $\nu(M,z)$, defined as the ratio between the critical overdensity
$\delta_{\rm crit}(z)$ for collapse at redshift $z$ and the linear rms fluctuation
at $z$ within spheres containing mass $M$. The larger the value of $\nu$ the
rarer and more massive the corresponding halo. Note that this scaling accounts
satisfactorily for the redshift dependence of the mass-concentration relation
shown in the left-hand panel.}
\label{fig:alphamz}
\end{figure*}

\subsection{Halo Catalogues}
\label{ssec:hcat}

Our halo cataloguing procedure starts from a standard $b=0.2$
friends-of-friends list of particle groups (Davis et al. 1985) and refines
it with the help of SUBFIND, the {\it subhalo finder} algorithm described by
Springel et al. (2001). Each halo is centred at the location of the minimum of
the gravitational potential of its main SUBFIND subhalo, and this centre is
used to compute the virial radius and mass of each halo.

We define the virial radius, $r_{\Delta}$, as that of a sphere of mean density
$\Delta \times \rho_{\rm crit}$. Note that this defines implicitly the
``virial mass'' of the halo, $M_{\Delta}$, as that enclosed within
$r_{\Delta}$.  The choice of $\Delta$ varies in the literature. The most
popular choices are: (i) a fixed value, as in NFW's original work, where
$\Delta=200$ was adopted; or (ii) a value motivated by the spherical top-hat
collapse model, where $\Delta\sim 178\, \Omega_{\rm m}^{0.45}$ for a flat
universe (Eke et al. 1996). The latter choice gives $\Delta=95.4$ at $z=0$ for
the cosmological parameters adopted for the {\tt MS}.  We keep track of both
definitions in our halo catalogue, and will specify our choice by a
subscript. Thus, $M_{200}$ and $r_{200}$ are the mass and radius of a halo
adopting $\Delta=200$, whereas $M_{\rm vir}$ and $r_{\rm vir}$ are the values
corresponding to $\Delta=95.4$ at $z=0$. Quantities specified without
subscript assume $\Delta=200$, so that, e.g., $M=M_{200}$. The halo
concentration is defined as the ratio between the virial radius and the scale
radius: $c=c_{200}=r_{200}/r_s$. In this case, the characteristic density is
related to the concentration by
\begin{equation}
\delta_c=(200/3)\, c^3/[\ln(1+c)-c/(1+c)].
\label{eq:cdeltac}
\end{equation}

Since halos are dynamically evolving structures, we use a combination of
diagnostics in order to flag non-equilibrium systems. Following
Neto et al. (2007), we assess the equilibrium state of each halo by measuring:
(i) the {\bf substructure mass fraction}; i.e., the total mass fraction in
resolved substructures whose centres lie within $r_{200}$, $f_{\rm
sub}=\sum_{i\neq0}^{N_{\rm sub}}M_{\rm sub,i}/M_{200}$; (ii) the {\bf centre
of mass displacement}, $s=|{\mathbf r}_{\rm c}-{\mathbf r}_{\rm cm}|/r_{200}$,
defined as the normalized offset between the location of the minimum of the
potential and the barycentre of the mass within $r_{200}$; and (iii) the {\bf
virial ratio}, $2T/|U|$, where $T$ is the total kinetic energy of the halo
particles within $r_{\rm vir}$ and $U$ their gravitational self-potential
energy.

Using these criteria, we shall consider halos to be {\it relaxed} if they
satisfy {\it all} the following conditions: $f_{\rm sub}<0.1$, $s<0.07$, and
$2T/|U|<1.35$. (See Neto et al. 2007 for full details.) 
We shall also impose a minimum number of particles in order to
be able to say something meaningful about internal halo structure. 
We initially set this limit at 500 particles, but following the analysis
of profile shapes presented in Section~3.2 we subsequently adopt a
stricter criterion of 3000 particles. We consider only {\it relaxed}
halos in this study, since only for such systems can the mass profiles
of individual objects be represented accurately by a simple fitting
formula with a small number of parameters. Such formulae are also
useful for characterizing the {\it average} profiles of large
ensembles of halos, since the fluctuations in the individual systems
then average out. Hayashi \& White (2007) present such mean profiles
as a function of mass for all {\tt MS} halos regardless of their
dynamical state.  We compare their results with our own below.

With these restrictions, including the 3000 particle limit, our final
{\tt MS} catalogue contains $128233$, $77190$, $30603$, and $9392$ relaxed
halos at $z=0$, $1$, $2$, and $3$, respectively. (The corresponding numbers
for the {\tt hMS} catalogue are $8131$, $6652$, $4112$ and $2194$.) The
overall fractions of these halos that are relaxed in the {\tt MS} are $78\%$,
$60\%$, $50\%$, and $48\%$ at these redshifts, respectively.  We note that
these fractions also depend on halo mass: at $z=0$ about $\sim 85\%$ of $\sim
10^{12} \, h^{-1} M_\odot$ halos are relaxed by our criteria, but only $\sim
60\%$ of $\sim 10^{15} \, h^{-1} M_\odot$ halos. In order to obtain usefully
large statistical samples, we restrict our analysis to the redshift range
$0\leq z\leq 3$ in the following.

\section{Halo density profiles}
\label{sec:res}

For each halo in the samples described in Section~\ref{sec:numexp} we have
computed a spherically-averaged density profile by measuring the halo mass in
$32$ equal intervals of $\log_{10}(r)$ over the range $0\geq
\log_{10}(r/r_{\rm vir})\geq -2.5$. 

Profiles may also be stacked in order to obtain an {\it average} profile for
halos of similar mass. This procedure has the advantage of erasing individual
deviations from a smooth profile which are typically due to the presence of
substructure. Such deviations increase the (already considerable) scatter in
the parameters fitted to individual profiles and may mask underlying trends in
the data.  The left panels in Fig.~\ref{fig:r2rho} show the profile that
results from stacking $464$ halos of mass $\sim 2\times 10^{14} h^{-1}
M_\odot$ identified at $z=0$ in the {\tt MS}. We choose to show $r^2 \rho$ vs
$r$ rather than $\rho$ vs $r$ in order to remove the main radial trend and
enhance the dynamic range of the plot. Similar stacked profiles for {\it all}
halos in the {\tt MS} (regardless of dynamical state) are shown by Hayashi
\& White (2008).

\subsection{Deviations from NFW and concentration estimates}

The concentration of a halo is defined using the scale radius of the profile,
$r_s$. This identifies the location of the maximum of the $r^2\rho$
profile. However, as shown in Fig.~\ref{fig:r2rho}, the peak is rather broad,
leading to some uncertainty in its exact location when noise is
present. Typically this problem is addressed by fitting the numerical data to
some specified functional form over an extended radial range.  The curves in
the top-left panel of Fig.~\ref{fig:r2rho} show the result of fitting the
stacked halo profile with the NFW formula (eq.~(\ref{eq:nfw})), but varying
the radial range of the fit as shown by the labels.  This results in slightly
different estimates for $r_s$, and consequently for the concentration, $c_{\rm
vir}=r_{\rm vir}/r_s$. Increasing the innermost radius of the fit from $0.02$
to $0.1 \,r_{\rm vir}$ results in a concentration estimate that decreases from
$\sim 7.5$ to $\sim 6.7$.

This variation is a result of the slight (but significant) mismatch between
the {\it shape} of the NFW profile and that of the stacked halo, as shown by
the residuals in the top-right panel of Fig.~\ref{fig:r2rho}.  The ``S'' shape
of the residuals implies that the stacked profile steepens more gradually with
radius in a log-log plot than does the NFW profile, a result that has been
discussed in detail by Navarro et al. (2004), Prada et al. (2006) and Merritt
et al. (2006).

These results suggest that force-fitting NFW profiles may induce spurious
correlations between mass and concentration. In particular, when halos are
identified in a single cosmological simulation, the numerical resolution
varies systematically with halo mass (less massive halos are more poorly
resolved) so that the radial range available for fitting is a strong function
of halo mass.

One way to address this issue is to adopt a fitting formula that better
matches the mean profile of simulated halos. As discussed by Navarro et
al. (2004), Prada et al. (2006) and Merritt et al. (2006), improved fits are
obtained adopting a radial density law where the logarithmic slope is a
power-law of radius,
\begin{equation}
{{d \log \rho} \over {d \log r}}= -2 \left({r \over r_{-2}}\right)^\alpha ,
\label{eq:dlnrhodlnr}
\end{equation}
which implies a density law of the form
\begin{equation}
\ln (\rho/\rho_{-2}) = -(2/\alpha) [(r/r_{-2})^\alpha -1].
\label{eq:rhoalpha}
\end{equation}
Here $\rho_{-2}$ is the density at $r=r_{-2}$. This density law was first
introduced by Einasto (1965) who used it to describe the distribution
of old stars within the Milky Way. For convenience, we will thus refer
to it as the Einasto profile. Note that according to our definition,
$r_{-2}$ corresponds to the radius where the logarithmic slope of the
density profile has the ``isothermal'' value, $-2$. In this sense,
$r_{-2}$ is equivalent to the NFW scale-length, $r_s$, and again
marks the location of the maximum in the $r^2 \rho$ profile shown in
Fig.~\ref{fig:r2rho}.

The bottom-left panel of Fig.~\ref{fig:r2rho} shows that, at the cost of
introducing an extra shape parameter, the fits improve to the point that the
sensitivity of concentration estimates to the fitted radial range is
effectively eliminated.  Thus, concentrations obtained by fitting
eq.~(\ref{eq:rhoalpha}) to the stacked halo profiles are robust against
variations in fitting details.  Hayashi \& White (2007) show that the same is
true for fits to stacks of {\it all} halos of a given mass, rather than just
the relaxed halos used to make Fig.~\ref{fig:r2rho}, and indeed, the
$\alpha$ and $c$ values they find are very similar to the values we obtain
here, showing that our restriction to relaxed halos has relatively little
effect in the mean.

We conclude that accounting for the subtle difference between halo profile
shape and the NFW fitting formula is worthwhile in order to avoid possible
fitting-induced biases in concentration estimates. In the remainder of this
paper we quote concentrations, $c_{\Delta}=r_{\Delta}/r_{-2}$, which are
estimated by fitting density profiles by eq.~(\ref{eq:rhoalpha}). We discuss
in the next section how $\alpha$ is chosen for these fits.

\begin{figure} \hspace{-0.5cm}
\resizebox{8cm}{!}{\includegraphics{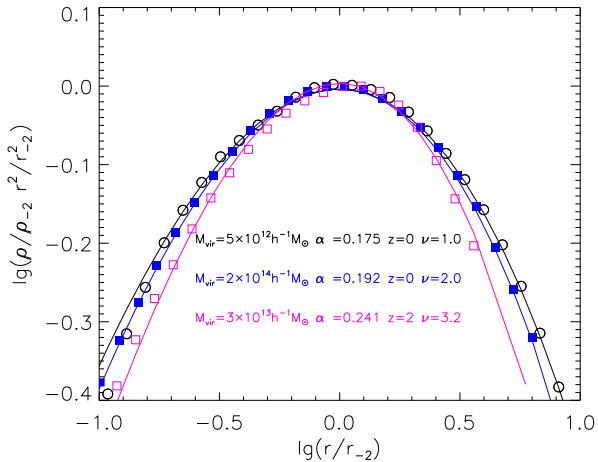}}
\caption{Three stacked halo density profiles with different values of $\nu$, and
at different redshifts, as labelled. The profiles are chosen to illustrate the
variation in halo structure as a function of $\nu$ indicated in Fig. 2. The
larger the halo mass, the larger the value of $\alpha$ and the sharper the
curvature in the density profile as a function of radius. }
\label{fig:r2rhonu}
\end{figure}

\begin{figure} \hspace{-0.5cm}
\resizebox{8cm}{!}{\includegraphics{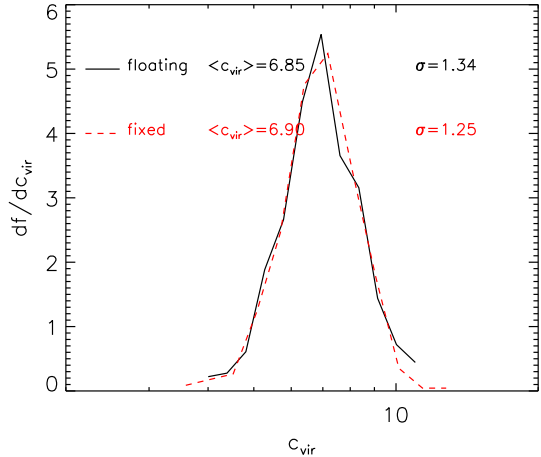}}
\caption{Distributions of concentration parameters estimated using
eq.~(\ref{eq:rhoalpha}) for the $464$ individual halo profiles stacked in
Fig.~\ref{fig:r2rho}. The solid (black) histogram corresponds to fits where
$\alpha$ was adjusted separately for each individual halo, the dashed (red)
histogram to fits where $\alpha$ was set to the value implied by the
$\alpha(\nu)$ relation of Fig.~\ref{fig:alphamz} (eq.~(\ref{eq:alphanu})). The
numbers in the legend give the median and scatter of the
distributions. The excellent agreement between the two distributions indicates
that unbiased and accurate concentration estimates for relaxed halos may be
obtained by fixing $\alpha$ to the value determined by eq.~(\ref{eq:alphanu}).
\label{fig:resid}}
\end{figure}

\subsection{The mass and redshift dependence of profile shape}
\label{ssec:smz}

The above discussion suggests that the shape parameter, $\alpha$, should be
used to improve the description of the typical density profiles of simulated
halos and to eliminate possible biases in estimates of their concentration.
To do this, it is necessary to understand whether (and how) $\alpha$ varies
with redshift and/or halo mass.

We explore this in Fig.~\ref{fig:alphamz}. The left panel shows how the
best-fit value of $\alpha$ depends on halo mass for the average profiles of
relaxed halos stacked according to their mass. We consider only halos with at
least $500$ particles within the virial radius, and stacks containing at
least $10$ halos. The solid and dashed curves in this plot correspond to the
{\tt MS} and {\tt hMS} simulations, respectively.

This figure illustrates several interesting points. In the first place, we note
that there is good agreement between the two simulations for halos which are
represented by at least 3000 particles, but that systematic
differences are visible when the {\tt MS} halos contain fewer particles than
this. In the rest of this paper we will thus only present results for halos
containing at least 3000 particles within the virial radius.

\begin{figure*} \hspace{-0.5cm}
\resizebox{16cm}{!}{\includegraphics{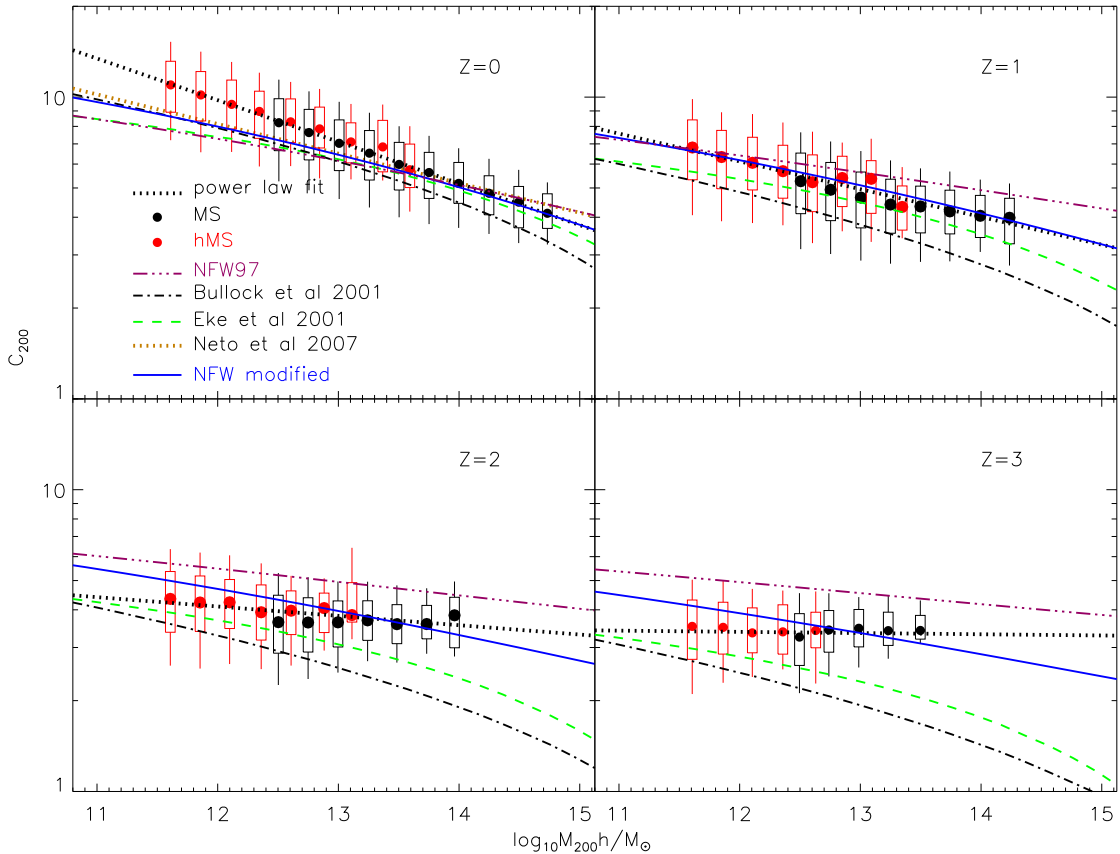}}
\caption{ The mass dependence of concentration as a function of
redshift. Concentration estimates are derived from Einasto fits to the density
profile of ``relaxed'' halos over the radial range $[0.05,1] r_{\rm vir}$.
Points, boxes and whiskers show the median and the 5, 25, 75, and 95
percentiles of the concentration distribution within each mass bin. Black
symbols correspond to the {\tt MS}, red symbols to the smaller, but higher
resolution {\tt hMS}. Only results for halos with more than $3000$ particles
are shown for each simulation. The solid-dotted, dashed, dot-dashed and solid
lines show the models of NFW, ENS, B01 and the modified NFW, respectively, for
comparison with our results. The lower dotted line in the upper left
panel indicates the relation obtained by Neto et al. (2007) 
from NFW fits to relaxed halos at $z=0$. The remaining dotted lines
show the powerlaw fits given in Table~1.}
\label{fig:cmz}
\end{figure*}

A second interesting point is that there are well-defined trends for $\alpha$
to increase with mass at each redshift, and with redshift at each mass. The
(weak) trend with mass was already visible in Navarro et al. (2004) and Merritt
et al. (2005), although the small number of halos in these studies made their
results rather inconclusive in the face of the large halo-to-halo scatter. The
use of stacked profiles in Fig.~\ref{fig:alphamz}, together with the much
larger number of halos in the simulations used here, leads to a far more
convincing demonstration of the trends than was previously possible.

The trend with redshift at given mass may seem surprising, but its
interpretation is made clear by the right panel of
Fig.~\ref{fig:alphamz}. Here we show $\alpha$ as a function of a dimensionless
``peak-height'' parameter, defined as $\nu(M,z) \equiv \delta_{\rm
crit}(z)/\sigma(M,z)$; i.e., as the ratio of the linear density threshold for
collapse at $z$ to the rms linear density fluctuation at $z$ within spheres of
mean enclosed mass $M$.  The parameter $\nu(M,z)$ is related to the abundance
of objects of mass $M$ at redshift $z$. $\nu(M_*,z)=1$ defines the
characteristic mass, $M_*(z)$, of the halo mass distribution at redshift
$z$. $\nu(M,z) \gg 1$ then corresponds to rare objects with $M \gg M_*$, while
$\nu(M,z) \ll 1$ corresponds to objects in the low-mass tail of the
distribution. The parameter $\nu$ plays a key role in the Press-Schechter
model for the growth of nonlinear structure (see, for example, Lacey \& Cole
1993).

The right panel of Fig.~\ref{fig:alphamz} shows that all curves coincide when
expressed in terms of $\nu(M,z)$ (and when considering halos with $N>3000$
particles). Thus, the redshift dependence in the left panel merely reflects
the fact that objects of given virial mass correspond to very different values
of $\nu$ at different redshifts. It is interesting that the dependence of
$\alpha$ on $\nu$ is very weak for $\nu<1$ (it is nearly constant at $\alpha
\sim 0.16$), but it increases rapidly for rarer, more massive objects,
reaching $\alpha \sim 0.3$ for $\nu \sim 3.5$. A simple formula,
\begin{equation}
\alpha=0.155 + 0.0095\, \nu^2
\label{eq:alphanu}
\end{equation}
describes the numerical results quite accurately.

Taylor \& Navarro (2001), Austin et al (2005) and Dehnen \& McLaughlin
(2005) investigated a halo model based on the assumption that the
phase space density, $\rho/\sigma^3$, was a simple powerlaw of
radius. Interestingly, the density profile they found is almost
identical to an Einasto profile with $0.14<\alpha<0.18$ and so the
sharp upturn we find in $\alpha$ at $\nu>1$ could be taken as
indication that such a model is not valid for rare and massive halos.

The dependence of profile shape on $\nu$ is illustrated in
Fig.~\ref{fig:r2rhonu}, where we plot $r^2 \rho$ profiles for halo stacks
corresponding to three different values of $\nu$: $1.0$, $2.0$, and $3.2$. The
larger $\nu$, the larger $\alpha$, and the more sharply the profile peaks.  It
is unclear what causes these systematic trends, but the fact that they depend
on $\nu$ (rather than, say, on mass alone) is an important clue for models
that attempt to explain the near-universality of dark halo density profiles.
While our results here are based purely on relaxed halos, very similar results
were found by Hayashi \& White (2007) for the average profiles of stacks of
{\it all} {\tt MS} halos, regardless of dynamical state.

\subsection{Concentration estimates}

As we discussed above, a fitting formula other than NFW is needed to obtain
concentration estimates that are insensitive to the radial range
fitted. Einasto's profile, eq.~(\ref{eq:rhoalpha}), accomplishes this at the
cost of introducing an additional shape parameter, $\alpha$. Adjusting
$\alpha$ to fit the detailed structure of individual halos would negate the
spirit of the NFW programme which attempts to predict the structure of dark
halos in any hierarchical cosmology from its initial power spectrum and the
global cosmological parameters. Three-parameter fits to individual halos are
also susceptible to strongly correlated parameter
errors. We have therefore explored whether fixing $\alpha$ as a function of
halo mass and redshift according to the $\alpha(\nu)$ relation of
eq.~(\ref{eq:alphanu}) results in significantly different concentration
distributions than adjusting it freely to fit each halo.

We show the result in Fig.~\ref{fig:resid}, which shows concentration
distributions for the $464$ individual halos which were stacked to make
Fig.~\ref{fig:r2rho}. We compare the distribution obtained from full
three-parameter Einasto fits to that obtained when $\alpha$ is set to the
value predicted for each halo from its mass. We find that for most halos the
fits give essentially the same concentration, and the distributions of
concentration are almost indistinguishable. The medians of the fixed-$\alpha$
and floating-$\alpha$ distributions are $6.85$ and $6.90$, respectively. 
The scatters also coincide, as noted in the labels of
Fig.~\ref{fig:resid}. Interestingly, the scatter around the best-fit profile
is typically only slightly smaller for the Einasto model than for the NFW
model, showing that the difference between the two models is much
smaller than the deviation of the profile of typical individual halos from
either model. In this sense, the 3-parameter Einasto profile has little
advantage over the 2-parameter NFW profile when estimating dark matter
halo concentrations.

This is easily understood. Estimating the concentration of a halo is
equivalent to locating the ``peak'' of the $r^2 \rho$ profile. Provided that
the shape of the fitted profile is, on average, a good approximation to the
simulated ones, no systematic offset is expected between concentrations
estimated using the two procedures. We conclude that concentrations may be
estimated robustly by fitting Einasto profiles to individual halos with
$\alpha$ fixed to the value obtained from eq.~(\ref{eq:alphanu}). All values
reported below were obtained using this procedure.

\subsection{The mass and redshift dependence of halo concentration}
\label{ssec:cmz}

The concentration-mass relation is shown in Fig.~\ref{fig:cmz} for
$z=0, 1, 2$ and $3$. The upper-left panel is equivalent to Fig.~6 of
Neto et al. (2007), except that our concentrations are estimated 
from fits covering the range $[0.05,1]r_{\rm vir}$  using an Einasto
rather than an NFW profile, and we only show results for halos with
more than $3000$ particles.  The differences are small, as may be
judged from the power-law fit proposed by Neto et al., $c_{200}=5.26
\, (M_{200}/10^{14} h^{-1} M_\odot)^{-0.10}$, shown as a dotted line
in the upper-left panel of Fig.~\ref{fig:cmz}. This power law fit is
very similar (after correction for differing concentration and mass
definitions) to that which Maccio et al. (2007) obtained from NFW fits
to halos in an independent set of (smaller) simulations, despite the
fact that these authors included halos with as few as 250 particles,
which we would consider to be significantly under-resolved on the
basis of our own tests.

Both the concentration values and the trends with mass and redshift are
very similar to those presented in figure~2 of Zhao et al (2003b) who
analysed a set of $\Lambda$CDM simulations of varying resolution. The small
offsets between their mean concentrations and our results are consistent
with the slight differences we have noted when switching from NFW to Einasto
models for determining concentrations. In addition to confirming these
earlier results, the much larger volume of our simulations results in a
better determination of the intrinsic scatter about the mean relation.

Red and black symbols in Fig.~\ref{fig:cmz} correspond to the {\tt MS}
and {\tt hMS} simulations, respectively, with dots plotted at the
median concentration for each mass bin. Boxes represent the lower and
upper quartiles of the concentration distributions, while the whiskers
show their 5\% and 95\% tails.  We also show the concentrations
predicted by three previously proposed analytic models: NFW
(solid-dotted magenta), B01 (dot-dashed black), ENS (dashed green). As
discussed by Neto et al. (2007), none of these models reproduces the
simulation results over the full mass range accessible at $z=0$: the
NFW and ENS models underestimate the concentration of galaxy-sized
halos, whereas the B01 model fails dramatically at the high-mass end,
where it predicts a sharp decline which is not seen in the simulations.
\footnote{We note that NFW, ENS and B01 parametrize the initial power spectrum
in slightly different ways, and that the predicted concentrations are
sensitive to the exact choice of parameters. For example, NFW and ENS use the
parameter $\Gamma$ to characterize the shape of the linear power spectrum. We
use $\Gamma=0.15$ here since that provides the best fit to the {\it actual}
power spectrum used in the {\tt MS}. For B01 we have used the default values
advocated in the latest version of their software ($K=2.8$ and $F=0.001$ in
their notation), which is available from {\tt
http://www.physics.uci.edu/$\sim$bullock/CVIR/}. The differences between the
predictions shown here and in Fig.~4 of Hayashi \& White (2008), or in Neto et
al (2007), for example, are due to slight differences in the values chosen for
these parameters.}

There is a hint in the $z=0$ panel that the relation is flattening at the
high-mass end, where the NFW predictions at $z=0$ appear slightly better than
those of ENS.  This is because a constant concentration for very rare and
massive objects is {\it implicit} in the NFW model, which assumes that the
characteristic density of a halo reflects that of the universe at the time it
collapsed.  Very massive systems assembled very recently, and therefore share
the same collapse time (i.e., they are being assembled today).

The near-constant concentration of the most massive halos is
considerably more obvious at higher redshift, presumably because
well-resolved halos (i.e., those with $N>3000$ in the {\tt MS} or {\tt
hMS}) become rarer and rarer with increasing lookback time. Indeed,
whereas at $z=0$ our {\tt MS} halo catalogue spans the range
$0.75<\nu<3$, at $z = 3$ all the halos retained have $\nu \, \gsim \,
3$. As a result, the average concentration at $z = 3$ is almost
independent of mass over the accessible mass range, i.e., $M\, \gsim
\, 3\times 10^{11} h^{-1} M_\odot$. It is interesting that the average
concentration of the most massive halos (i.e. $\nu \geq 3$) is similar
at all redshifts, $c\sim 3.5$ to $4$. This evokes the proposals of
Zhao et al. (2003a,b), Tasitsiomi et al. (2004) and Lu et al. (2006),
all of whom argue that halos undergoing rapid growth should all have
similar concentration.


The evolution of the mass-concentration relation seen in our numerical
simulations is not predicted by any of the published prescriptions.  The
original NFW model shows a flattening of the concentration-mass relation with
increasing redshift, similar to that observed in the simulations, but it
predicts insufficient evolution in concentration at given mass. As a result,
this model overestimates the concentrations by an increasing amount with
increasing redshift, about 40\% at $z=3$. The ENS and B01 models fail to
reproduce the features of the mass-concentration-redshift relations at high
mass, predicting a stronger mass dependence and much more evolution than is
seen. In these two models, the concentration of halos of given mass scales
inversely as the expansion factor, so that shape of the mass-concentration
relation remains fixed. While the high-mass discrepancy between the B01
model and our measurements can be reduced by changing the parameters to
$K=2.8$ and $F=0.001$ (see Wechsler et al 2006), neither for this model nor
for the ENS model can parameter changes produce agreement with the weak
redshift evolution seen at high mass both here and by Zhao et al. (2003b).  On
the other hand, the ENS and B01 models predict the concentration evolution of
galaxy mass halos substantially better than the NFW model. Note that our
simulation data do not disagree significantly from those analysed by ENS and
B01. The discrepancies result from extrapolation of their proposed relations
beyond the range where they were reliably tested.

\begin{table}
\label{tab_fits}
\caption{Values of the constants $A$ and $B$ for the best straight-line fit
(\ref{eq:best-fit}) to the data shown in Fig.~\ref{fig:cmz}.}  \centering
\begin{tabular}{ccc}
\hline\hline
Redshift & $A$ & $B$ \\
\hline
0   & -0.138 & 2.646 \\
0.5 & -0.125 & 2.372 \\
1   & -0.092 & 1.891 \\
2   & -0.031 & 0.985 \\
3   & -0.004 & 0.577 \\
\end{tabular}
\label{tab:best-fit}
\end{table}

In the NFW model, the definition of formation time involves two
physical parameters, $F$ and $f$, and a proportionality constant, $C$,
that relates the value of the characteristic halo density to the mean
density of the universe at the time of collapse. The formation
redshift of a halo is taken to be the redshift at which a fraction,
$F$, of its mass was first contained in progenitors each individually
containing at least a fraction $f$ of its mass. In the original NFW
prescription, $F=0.5$, $f=0.01$ and $C=3000$. We find that the
observed trend in the slope of the mass-concentration relation with
redshift can be approximately reproduced by simply changing $F$ to
$F=0.1$, keeping $f=0.01$ as before; the normalization of the relation
is then approximately reproduced by taking $C=600$. The resulting curves
are shown as solid blue lines in Fig.~\ref{fig:cmz}. This modified NFW model
succeeds well in matching the redshift evolution, but its mass 
dependence is still too shallow at $z=0$ and too steep at $z=3$,
leaving room for improvement, especially at low masses where the B01 and ENS
prescriptions give better predictions of the evolution rate. We have
checked that the same model also gives an acceptable fit for other
cosmologies, in particular for a simulation of a $\Lambda$CDM model
similar to {\tt hMS } but with the values of the cosmological
parameters inferred from the 3-year WMAP satellite data (Spergel et
al. 2007) ($\Omega=0.236$, $\Omega_{\Lambda}=0.764$, $n=0.97$ and
$\sigma_8=0.74$). We have also checked that this modified NFW model
gives a good description of the concentrations found in the scale-free
models presented in Navarro, Frenk \& White (1997) which span a range
of spectral indices and have either $\Omega=1$ or $\Omega=0.1$. If
more accurate results for the particular cosmology assumed in this
paper are desired, we provide the coefficients of power-law fits of the form
\begin{equation}
{\rm log}_{10}(c_{200}) = A{\rm log}_{10}(M_{200}) + B
\label{eq:best-fit}
\end{equation}
in Table~\ref{tab:best-fit} and plot these
best fit curves in Fig.~\ref{fig:cmz}.

The disagreement between the simulation data and the original NFW
prescription is up to a few tens of percent. The BO1 and ENS
prescriptions underestimate the concentration of the most massive
halos by factors of $\sim 2$ to $\sim 3$ at high redshift. Since the
characteristic density of a halo scales roughly as the third power of
the concentration (see eq.~(\ref{eq:cdeltac})), this implies that the
characteristic densities of such halos are underestimated by at least
an order of magnitude by the latter two models, leading to dramatic
changes in the expected lensing power, X-ray luminosity, and S-Z
detectability of massive clusters at high redshift.

It is also interesting to compare the $z=0$ concentration-mass
relation found here to the one which Hayashi \& White (2007) obtained
by fitting Einasto profiles to stacks of {\it all} halos of a given
mass, rather than to stacks of relaxed halos. The results in their
Fig.~4 lie very close to the NFW relation plotted in the upper-left
panel of our own Fig.~\ref{fig:cmz}. Thus, the restriction to relaxed
halos has little systematic effect on Einasto-based concentrations at
masses above about $3\times 10^{13}h^{-1}M_\odot$, but at lower masses
it results in somewhat larger concentrations. This differs slightly
from the result of Neto et al (2007) who found $z=0$ concentrations
obtained by fitting NFW (rather than Einasto) profiles to relaxed
halos to be greater by a larger amount at high mass.  The median of
their NFW-based concentrations is also closer to the NFW model
prediction on galaxy scales (both for relaxed and for all halos) than
is the median of the Einasto-based concentrations which we plot in
Fig.~\ref{fig:cmz}. This is at least in part a result of the bias
illustrated in Fig.~\ref{fig:r2rho}.

Neto et al. (2007), Hayashi \& White (2007) and this paper are all based on
the same simulations (although Hayashi \& White do not use the {\tt hMS}).
The small but (statistically) significant differences in their derived
concentration-mass relations show that at the 10 to 20\% level these relations
depend on the details of the halo sample selection and profile fitting
procedures.  The {\bf modified} NFW model works reasonably well over
the full range of mass and redshift studied here, certainly rather
better than the revisions proposed by ENS and B01. Significant
discrepancies remain, however, notably at galaxy masses where the NFW model
underpredicts the rate of evolution, resulting in systematically low
Einasto-based concentrations at $z\sim 0$ and systematically high values at
$z\geq 2$.  For such objects the evolution {\it rate} is better predicted by the ENS
and B01 models. Given the good agreement between various simulations on this
mass scale (see, e.g., Zhao et al 2003b, Macci\`o et al. 2007 and Neto
et al. 2007), we believe these concentration estimates for low redshift
$\Lambda$CDM galaxy halos to be accurate. This implies that the NFW model
needs further improvement for such objects. At higher masses and higher
redshifts, however, the (modified) NFW formalism works quite well, and should
be preferred to those of B01 or ENS.

With our modified parameters, the NFW prescription not only fits
concentrations in the WMAP1 cosmology investigated here, but also our
(less extensive) set of data for the WMAP3 cosmology.  Without
detailed testing, however, it is unclear if the prescription can be
extended to other regimes of interest. For example, what
concentrations are expected for dwarf galaxy halos or for ``first
object'' halos at $z \sim 30$?  The incorrect conclusions reached by
applying the B01 and ENS formulae to massive clusters at high redshift
are testimony to the dangers of using empirical formulae outside the
range where they have been tested. Clearly, further theoretical effort
aimed at understanding the factors which determine the internal
structure of dark halos would be a welcome complement to the numerical
work presented here.

\section{Summary}
\label{sec:conc}

We have used data from the {\it Millennium Simulation} and from a smaller but
higher resolution simulation to examine how the density profiles of relaxed
$\Lambda$CDM halos vary with halo mass and redshift. We study profile shape
and concentration for halos with mass exceeding $\sim 3 \times 10^{11} h^{-1}
M_\odot$ over the redshift range $0\leq z\leq 3$.  Our main results may be
summarized as follows.

\begin{itemize}

\item We confirm the conclusion of previous studies that, in the mean, the
shape of spherically averaged $\Lambda$CDM halo density profiles deviates
slightly but systematically from the two-parameter fitting formula proposed by
Navarro, Frenk \& White (1995, 1996, 1997). A more accurate description is
provided by the three-parameter Einasto (1965) profile, for which the
logarithmic slope is a power-law of radius, $d\log \rho/d\log r \propto
r^{\alpha}$. We show that this fitting formula gives concentration estimates
which are insensitive to the radial range fitted, albeit at the price of an
additional shape parameter.  Although Einasto fits avoid small but significant
biases that arise if all halos are fit to the NFW model, we emphasise that
individual halos typically deviate from either model by more than the
difference between them.

\item Using stacked profiles of halos of similar mass, we show that the shape
parameter, $\alpha$, depends systematically on halo mass and on redshift.
These dependences can be collapsed into a dependence on the single ``peak
height'' parameter, $\nu(M,z)=\delta_{\rm crit}(z)/\sigma(M,z)$.  Halos with
large $\nu$ are rare objects in the high-mass tail of the halo mass
distribution and have large $\alpha$ values (see eq.~(\ref{eq:alphanu})). This
provides an important clue for models that attempt to interpret the dependence
of halo density profiles on mass and redshift.

\item The dependence of halo concentration on mass becomes progressively
weaker with increasing redshift, as found earlier by Zhao et al
(2003b).  At $z = 3$, concentrations are almost independent of mass over the
mass range accessible to our simulations.  Relaxed halos with $\nu > 3$ (the
rarest and most massive systems) have similar concentrations, $\langle c_{200}
\rangle = 3.5$ to 4, at all the redshifts we have studied .

\item The models of Bullock et al. (2001) and Eke et al. (2001) fail to
reproduce our measured concentrations for high-mass and high-redshift objects,
predicting a stronger mass dependence and more evolution than is seen in the
simulations. Parameters in these models can be changed to reduce the
strength of their mass dependence, but the predicted redshift evolution,
while fitting galaxy mass halos well up to redshift $z=1$, remains
substantially too strong at high mass and at higher redshifts. As a
result, the predictions of these models for high-redshift galaxy
clusters can be in error by large factors. The original model of Navarro et
al. (1997) overpredicts the concentrations of such objects at redshifts beyond
1 (by up to $\sim 40\%$ at $z=3$) but a modified NFW model with a different
definition of formation redshift reproduces the simulation results
substantially better over the redshift and mass ranges we have examined. Both
the original and the modified NFW models underestimate the concentration
evolution of relaxed $10^{12}h^{-1}M_\odot$ halos, leading to 30\%
discrepancies at $z=0$ and $z=3$.

\end{itemize}

We hope that our simulation results will stimulate theoretical work aimed at a
deeper understanding of the factors which determine the internal structure of
$\Lambda$CDM halos. Such work may eventually result in simple recipes like
those of NFW, ENS and B01. Substantial errors are found, however, when these
published prescriptions are extrapolated beyond the regime where their authors
tested them, in particular, to the regime relevant to high-redshift galaxy
clusters. This demonstrates that careful numerical work is mandatory before
any recipe can be applied in a new regime. When making forecasts for surveys of
distant massive clusters, our simulations show our modified NFW recipe to give
reliable results at least out to $z=3$.

\section*{acknowledgments}

The {\it Millennium Simulation} was carried out as part of the programme of
the Virgo Consortium on the Regatta supercomputer of the Computing Centre of
the Max-Planck Society in Garching. The {\tt hMS} simulation was carried out
using the Cosmology Machine at Durham. We thank Adam Mantz for pointing out an
error in the NFW97 curve plotted in the original preprint version of Figure~5.
JFN acknowledges support from the Alexander von Humboldt Foundation and from
the Leverhulme Trust, as well as the hospitality of the Max-Planck Institute
for Astrophysics in Garching, Germany, and the Institute for Computational
Cosmology in Durham, UK. CSF ackowledges a Royal Society Wolfson Research
Merit Award. This work was supported in part by the PPARC Rolling Grant for
Extragalactic Astronomy and Cosmology at Durham.

\newpage
\appendix

\renewcommand{\theequation}{A\arabic{equation}}
\setcounter{equation}{0}  

\renewcommand{\thefigure}{A\arabic{figure}}
\setcounter{figure}{0}  
\end{document}